\documentclass[onecolumn,nofootinbib]{revtex4}

\usepackage{graphicx,epsfig}
\usepackage{multirow}
\usepackage{amsmath}
\usepackage{amsfonts}
\usepackage{fancyhdr}
\usepackage{hyperref}
\usepackage{xcolor}
\usepackage{float}
\hypersetup{colorlinks,linkcolor={blue},citecolor={blue},urlcolor={blue}}

\newcommand{\R}{{\cal R}}
\begin{document}

\pagestyle{fancy}
\lhead{\bf }
\rhead{}
\lfoot{}
\rfoot{}

\title{The trichotomy of primordial black holes initial conditions}
\author{Cristiano Germani}
\email[]{germani@icc.ub.edu}
\author{Laia Montell\`a}
\email[]{lmontella@icc.ub.edu}
\affiliation{Departament de F\'isica Qu\`antica i Astrof\'isica and Institut de Ci\`encies del Cosmos, Universitat de Barcelona, Mart\'i i Franqu\`es 1, 08028 Barcelona, Spain}

\begin{abstract}
We show that the threshold to form a black hole, in an asymptotically flat and radiation-dominated Friedman-Robertson-Walker (FRW) Universe, is not solely (mainly) determined by the behaviour of the compaction function at its extrema, as earlier thought, but also by the Ricci scalar of the spatial geometry at smaller (but super-horizon) scales, which we call ``the core''. We introduce three classes of initial conditions characterised by an open (O), closed (C), or flat (F) FRW core surrounded by a shell with higher three-dimensional curvature. In the C case, the core helps the collapse so that the black hole formation threshold is the lowest among all cases. Type-II black holes might only be generated by Type-O or F (each of those with different thresholds, with O being the highest) or by a Type-C with a negligible Ricci scalar at the centre, which we call an effective F core. Finally, we argue that an F core is typically more probable for a sharp power spectrum, however, it is also more likely related to non-spherical initial conditions. On the other hand, a very broad power spectrum, which might be related to the observed NanoGrav signal, would favour the formation of Type-I black holes with a mass spectrum peaked at the Infra-Red scale. 
\end{abstract}
\maketitle

\section{Introduction}
The knowledge of the threshold leading to black hole formation is a key ingredient in predicting their abundance in our Universe. In a seminal paper \cite{ss}, Shibata and Sasaki argued that the conditions to form a black hole from a large over-density is fully determined by the behaviour of the compaction function, which is basically the gravitational potential within a sphere of radius $r$, once the infinitely long wavelength part of the perturbation is extracted. In the case of Type-I black holes, where the compaction function has only one maximum, Musco \cite{musco} set the threshold to trigger gravitational collapse to the magnitude of this value. Although this threshold does depend upon the full shape of the compaction function \cite{musco}, it has been argued in \cite{universal} that it is mainly determined by its curvature at the maximum. With this at hand, an analytical formula, matching up to a few percent error, the numerical results \cite{albert}, has been provided. This opened up the doors to construct the non-linear statistics of primordial, Type-I, black holes \cite{sheth}. 

Type-II black holes instead, are characterised by the presence of a minimum in the compaction function, surrounded by two maxima. The distinction between this and the Type-I case is mathematically determined by the magnitude at the maximum ($r_m$) of a variable constructed from the {\it linear} smoothed over-density multiplied by the square of the background areal radius. This variable, dubbed as $g$, gained a central role in the statistics of Primordial Black Holes (PBHs) abundances \cite{sheth}. Statistically, $g$ inherits the probability distribution of curvature perturbations.

Until very recently, Type-II black holes were discarded as they needed larger values of $g$ ($>\frac{4}{3}$) than those of Type-I, and this was thought to be unlikely. Nevertheless, in a recent paper \cite{jacopo}, the Authors showed that, in the case of a broad power spectrum, the non-linear statistics of Type-I profiles, developed by the use of the analytical formula of \cite{universal}, prefer threshold values extremely close to Type-II amplitudes. 

Shortly after the breakthrough on numerical simulation of Type-II black holes \cite{T2}, which resolved the coordinate singularity found in \cite{kopp}, Escriv\`a in \cite{w}, showed that for certain curvature profiles of the linear smoothed over-density, parametrised by their dimensionless curvature at $r_m$ ($w$), would produce Type-I or II black holes dependently upon the size of $w$. This shed doubts on the validity of the analytical formula \cite{universal}, which, for Type-I, was supposed to be valid for any value of $w$.

What we are going to argue here is that there is a trichotomy of possible initial conditions that provides three different behaviours for the PBH formation thresholds: \\
Consider an isolated spherically symmetric scalar perturbation. At super-horizon scales, a local patch of the perturbed Universe can be approximated by a Friedman-Robertson-Walker (FRW) metric with certain constant three-dimensional curvature (separate Universe). This patch, which we call the {\it core}, is either a closed (Type-C), an open (Type-O) or a flat (Type-F) FRW Universe. By looking at larger distances, however, the spacetime becomes more and more inhomogeneous until reaching a peak of the co-moving three-dimensional curvature 
\begin{equation}
{\cal R}\equiv a^2 \times {}^{(3)}R\ ,
\end{equation}
which in turn is related to a shell with the highest non-linear over-density. This over-dense shell is the one that, if over-threshold, eventually collapses forming a black hole. 

What we empirically found is that, while Type-O and F would not follow the analytical formula of \cite{universal} for large values of $w$, as already noted in \cite{w} for the F case, while a Type-C fluctuation, with a non-negligible core, would instead follow it increasingly better for higher values of $w$. 

In other words, the three core cases provide three distinct thresholds for the maximum of $g$, with the C type threshold being the lowest. The reason of it is that in the C case, the core geometry would help the collapse. Whether this extra ingredient is relevant or not depends on the spatial extension and amplitude of the internal core, which is related a) to the sharpness of the surrounding shell parametrised by $w$ and b) by its amplitude: \\
For small values of $w$, the external shell is broad and overlaps the core so that the core loses importance. There, all three types of initial conditions follow reasonably well the analytical formula of \cite{universal} (see e.g. \cite{w} for the F case). On the contrary, for large $w$, the three thresholds would greatly differ from each other, as we shall show. Regarding the amplitude of the core, we found that Type-I is always obtained from Type-C if the ratio of the core curvature with respect to the peak one is higher than a threshold parametrised by $w_{\cal R}$, the dimensionless peak curvature of $\cal R$. Consistently, we find that for $w_{\cal R}\rightarrow 0$ ($\equiv w\rightarrow 0$) black holes are always of Type-I. On the other hand, for large $w_{\cal R}$, the threshold seems to saturate, and the formation of Type-I black holes is only guaranteed for core three-curvature amplitudes larger than $\sim 70\%$ of their peak value.

\section{Types of initial conditions}
Working beyond the separate universe approach, within gradient expansion, a spherically symmetric metric at super horizon scales\footnote{i.e. for wavelengths larger than $(a H)^{-1}$, where $a$ is the scale factor of the background FRW Universe and $H\equiv \frac{\dot{a}}{a}$.} can be written as
\begin{equation}\label{metric}
    ds^2=-dt^2+a^2(t)e^{2\zeta(r)}\left(dr^2+r^2d\Omega^2\right),
\end{equation}
or, equivalently
\begin{equation}\label{K}
    ds^2=-dt^2+a^2(t)\left(\frac{dx^2}{1-K(x)x^2}+x^2d\Omega^2\right).
\end{equation}

Using the first set of coordinates, we can write the compaction function \cite{ss}, $\mathcal{C}(r)$, in terms of the curvature fluctuation \cite{yoo} (in units of $G_N=1$)
\begin{equation}\label{comp}
    \mathcal{C}(r)=\frac{2}{3}\left[ 1-\left(1+r\zeta'(r)\right)^2\right]=g(r)\left(1-\frac{3}{8}g(r)\right),
\end{equation}
where we have defined, as in \cite{sheth}, $g(r)=-\frac{4}{3}r\zeta'(r)$, $r=0$ is the locus of the over-density peak (for a maximum) or trough (for a minimum). Mathematically, this variable can also be written in terms of the smoothed {\it linear} over density up to a radius $r$ ($\delta_r$) \cite{sheth}, i.e. $g\equiv a_i H_i^2 r^2 \delta_r$, where $H_i$ and $a_i$ (set to be $1$), are the initial Hubble and scale factors. The compaction function represents the gravitational potential enclosed in a radius $r$ generated by the non-linear over-density when the background energy-density is extracted. 

The maximum of $g$, which we shall name $r_m$, corresponds to a maximum of the compaction function for  $g(r_m)<\frac{4}{3}$ (Type-I black holes), and a minimum for $g(r_m)>\frac{4}{3}$ (Type-II). For Type-I black holes, the threshold is typically given in terms of the maximum of the compaction function, as there, the gravitational potential is the highest. In Type-II black holes, however, the minimum of the compaction function is surrounded by two maxima. The compaction function within the two maximums, being under-critical, fights against collapse, thus, the formation of a black hole is triggered whenever the magnitude of the non-linear over-density at $r=r_m$ is {\it smaller} than a threshold. In terms of the linear smoothed over-density at $r_m$, however, both cases would require $g(r_m)>g_c$, where $g_c$ is profile dependent \cite{musco}. 

Nevertheless, for Type-I black holes, it has been found in \cite{universal} that $g_c$ mostly depends on the physics around the collapsing shell surrounding the largest over-density value (the {\it shell} from now on). Specifically, the threshold has been found to depend upon the curvature of the smoothed linear over-density at its maximum, i.e., it would only depend on   
\begin{eqnarray}
    w\equiv-r^2\frac{d^2 g(r)}{dr^2}\Big|_{r=r_m}\ ,
\end{eqnarray}
which is its dimensionless curvature at $r_m$.

On the contrary, and this is what we are going to argue, in Type-II configurations, the central value of the over-density plays a non-trivial role for the threshold.
\subsection{The core}
According to gradient expansion, at super-horizon scales, a local patch of the spacetime is well approximated by a FRW Universe with a specific three-dimensional curvature, and the initial conditions for PBH formation can be classified by the sign of ${\cal R}(x\rightarrow 0)\sim 6K(x\rightarrow 0)$, as in the following table:
\begin{table}[H]
    \centering
    \begin{tabular}{|c|c|c|} \hline
        $\R(x\rightarrow 0)$ & Type & ``small scales'' FRW Universe \\ \hline\hline
         + & C & Closed  \\ \hline
         0 & F & Flat \\ \hline
         - & O & Open\\ \hline
    \end{tabular}
    \label{classification}
\end{table}
Primordial black holes are then formed by the next-to-leading order inhomogeneities in the three-dimensional curvature.

Because we shall mainly work with the metric \eqref{metric}, we need the following relation between $x$ and $r$:
\begin{equation}
    x=re^{\zeta(r)}.\label{xr}
\end{equation} 

For Type-II profiles, \eqref{xr} cannot be inverted as it passes through an extrema. Thus, we consider instead
\begin{equation}
    \R(r)=-\frac{2 e^{-2 \zeta (r)} \left(2 r \zeta ''(r)+r \zeta'(r)^2+4\zeta'(r)\right)}{r}.\label{Rr}
\end{equation}

Note that $\R$ is invariant under spatial coordinate transformations.

As we can already anticipate, the amplitude and size of the core can either help or resist the collapse of the shell. It indeed turns out, as we are going to see, that Type-II configurations are related to the case in which the core cannot be forgotten in the computation of the threshold.

\subsection{Initial conditions}
The profiles considered in \cite{w}, which for large values of $w$ implied Type-II configurations, correspond to Type-F. To make our point, we have taken the exponential (Type-F) profile as a base and then deformed it in order to obtain a closed or open FRW core. In this sense, what we are going to consider are collapsing shells characterised by the three possible configurations for the cores and connect those with Type-I and II black holes.

Starting from the base, we then consider: 
\begin{itemize}
\item Type-F:
\begin{equation}
    \zeta_F(r)=\mu\ e^{-(\frac{r}{r_m})^\beta}.\label{typeF}
\end{equation}
The parameter $\beta$ in \eqref{typeF} is related to the dimensionless curvature of $g(r)$ at $r_m$ ($w$).

It is easy to see that in $r\rightarrow 0$, $\R\rightarrow 0$
\begin{equation}
    \R(r)\approx  4\mu\ \beta(1+\beta)\frac{e^{-2\mu} }{r_m^\beta}r^{\beta-2}+\mathcal{O}(r^{2\beta-2}),
\end{equation}
valid for $\beta\ge 2$. This is a physical condition, as for smaller $\beta$ the Poisson equation would diverge at the centre.

\item Type-C: 
\begin{equation}
    \zeta_C(r)=\mu \left(1-\frac{\left(\lambda\frac{r}{r_m}\right)^2}{\alpha\left[1+\left(\lambda\frac{r}{r_m}\right)^3\right]}\right)e^{-(\lambda\frac{r}{r_m})^\beta},\label{typeI}
\end{equation}
where $\lambda$ is the solution of
\begin{equation}
    \left.\frac{dg(r)}{dr}\right|_{r=r_m}\propto\left.\zeta'(r)+r\zeta''(r) \right|_{r=r_m}=0,
\end{equation}
enforcing that $r_m$ is the maximum of $g(r)$.

The expansion of \eqref{Rr} for this profile corresponds to a closed FRW Universe:
\begin{equation}
    \R(r)\approx 24\mu  \frac{e^{-2\mu}\lambda^2}{\alpha r_m^2}+\mathcal{O}(r^{\beta-2}),\label{Rclose}
\end{equation}
valid for $\beta\ge 2$.

As we can see from Eq.\eqref{Rclose}, the role of the $\alpha$ parameter is to suppress the core amplitude of the perturbations. 

\item Type-O:
\begin{equation}
    \zeta_{O}(r)=\mu \left(1+\frac{\left(\lambda\frac{r}{r_m}\right)^2}{\alpha\left[1+\left(\lambda\frac{r}{r_m}\right)^3\right]}\right)e^{-\left(\lambda\frac{r}{r_m}\right)^\beta}.\label{TypeO}
\end{equation}
The expansion of $\R$ in this case,
\begin{equation}
    \R(r)\approx -24\mu  \frac{e^{-2\mu}\lambda^2}{\alpha r_m^2}+\mathcal{O}(r^{\beta-2}),
\end{equation}
is valid for $\beta\ge 2$.
\end{itemize}

\begin{figure}[h!]
  \centering
  \includegraphics[width=0.58\linewidth]{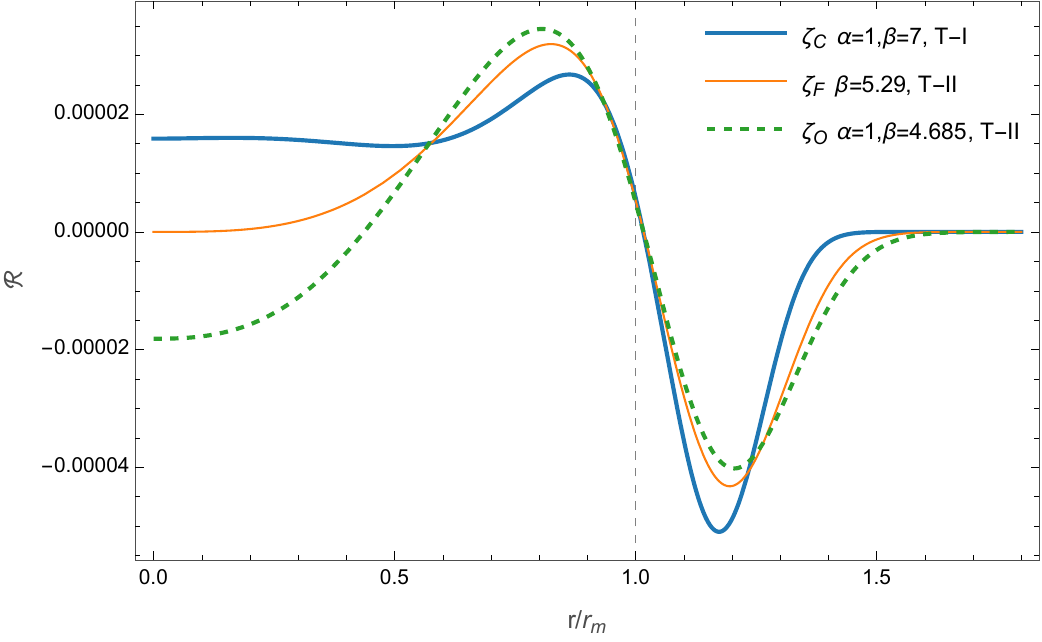}
  \caption{Typical $\R(r)$ for the three types of perturbation at threshold. All of them correspond to $w\simeq38.6$. The dashed vertical line is $r=r_m$.}
    \label{fig:Rall}
\end{figure}

Apart from the core behaviour, which is characteristic of each profile, all three profiles present a maximum of $\cal R$ (or equivalently the non-linear over-density) located at $r=r_p$ and a subsequent minimum, as can be seen in Fig.\ref{fig:Rall}. The radius $r_m$, discussed earlier, sits in between (see e.g. Fig.\ref{fig:Rall}). The similarities of the curves are due to the fact that we are considering only a single peak $\zeta$ profiles, the most luckily configuration for PBHs. 

Note that the threshold was historically given in terms of $\cal C$ and $g$, while the profiles and statistics in terms of $\zeta$. What we are going to argue is that the three-dimensional curvature, $\cal R$, is what it should be instead used.
\section{Numerics}

To check whether an initial perturbation collapses or not, and thus find the threshold in terms of $g(r_m)$, we have used the public SPriBHoS-II code \cite{codeII}. The threshold would initially be given in terms of a critical $\mu=\mu_c$ and then translated in terms of $g(r_m)$. 

To track the black hole formation more efficiently, we have modified the code, including an auxiliary function defined as twice the Misner-Sharp mass $M$ \cite{ms} over the areal radius $R$
\begin{equation}
    \mathcal{C}_0=2\frac{M}{R}\ .
\end{equation}
An apparent horizon leading to a black hole is only formed when $\mathcal{C}_0$, at its maximum, reaches $1$ twice. The first $1$ would signal the cosmological horizon crossing of the super-horizon perturbation. After this, because of the dilution of the Universe expansion, $\mathcal{C}_0$ would decrease. However, over threshold, the decay would then be followed by a growth, and a sub-horizon trapped surface would form whenever $\mathcal{C}_0\sim 1$ for the second time.

Large values of $w$ ($>30$), i.e. initial sharp profiles, required a substantial modification of the number of Chebyshev points ($N_{\text{cheb}}$) in the pseudo-spectral method used in \cite{w}. The code becomes increasingly unstable for $w>50$ in the Type-C case. However, as we shall show, the threshold already saturates at those values with no need of extra probes\footnote{In fact, Type-C configurations for $w>50$ when simulated with the SPriBHoS-II code, show a clear trend to the formation of a trapped horizon with a saturated threshold. This happens way before the simulation breaks because of Gibbs phenomena.}.

Finally, simulations are done in units $G_N=1$ with an initial Hubble scale $H_i=\frac{1}{2}$. All our numerical results will then be given in those units.

See Appendix \ref{appendix:hamiltonian} for the evolution of the Hamiltonian constraint.
\section{Thresholds} 
\begin{figure}[h!]
    \centering
    \includegraphics[width=0.6\linewidth]{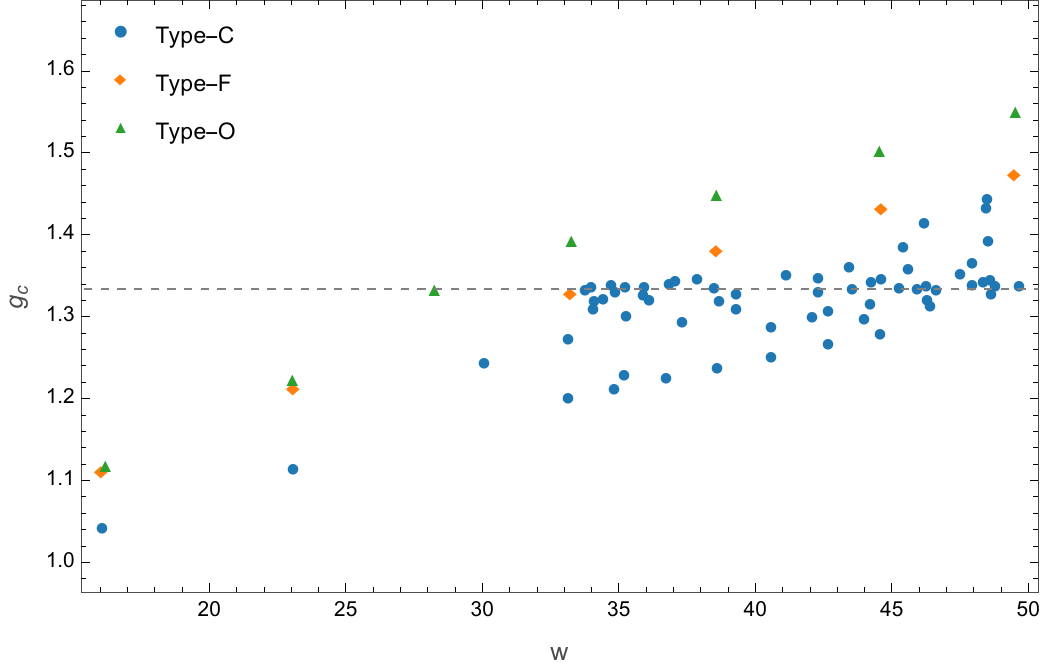}
    \caption{Numerical threshold, in terms of $g$, for the three types of perturbations. The dashed line corresponds to $g=\frac{4}{3}.$ All the points correspond to different configurations of the $\alpha$ and $\beta$ parameters (see Appendix \ref{appendix:parameters}).}
    \label{fig:gcAll}
\end{figure}

Here we present the threshold values, $g_c$, for different profiles. In Fig.\ref{fig:gcAll}, we can clearly see that as $w$ increases, the trichotomy appears. For Type-F and Type-C perturbations, the crossing between Type-I and Type-II happens around $w\sim34$, while for Type-O happens before, at $w\sim29$. We also see the trend anticipated earlier, that the lowest thresholds are related to Type-C, where the cores help the collapse. On the other hand, the highest thresholds are related to Type-O, where instead the cores fight the collapse. 

This is one of the main results of our paper, namely that for large enough cores, the threshold cannot only depend upon the shell structure.

For large values of $w$ (still in Fig. \ref{fig:gcAll}), Type-O and Type-F become Type-II, as already noted in \cite{w}. On the other hand, Type-C perturbations lie in both Type-I and Type-II regions. We will explore this peculiar behaviour in the following sections, nevertheless, we can already anticipate that Type-C should smoothly behave as Type-F for small core amplitudes. This is precisely what happens in the Fig. \ref{fig:gcAll}.

\subsection{Type-I}
The threshold for Type-I perturbations suggested by \cite{universal} was computed in terms of the dimensionless curvature of the compaction function ($q$)
\begin{equation}
    q=\frac{-r_m^2\partial_r^2\mathcal{C}(r_m)}{4\mathcal{C}(r_m)\left(1-\frac{3}{2}\mathcal{C}(r_m)\right)}.
\end{equation}
In terms of $\cal C$ at its maximum, it was then found to be 
\begin{equation}
    \mathcal{C}_c(q)=\frac{4}{15}e^{-\frac{1}{q}}\frac{q^{1-\frac{5}{2q}}}{\Gamma(\frac{5}{2q})-\Gamma(\frac{5}{2q},\frac{1}{q})}.\label{deltaA}
\end{equation}
which matches within a few percent error the threshold found in numerical simulations.

\begin{figure}[h!]
  \centering
  \begin{minipage}[c]{0.49\linewidth}
    \centering
    \includegraphics[width=\linewidth]{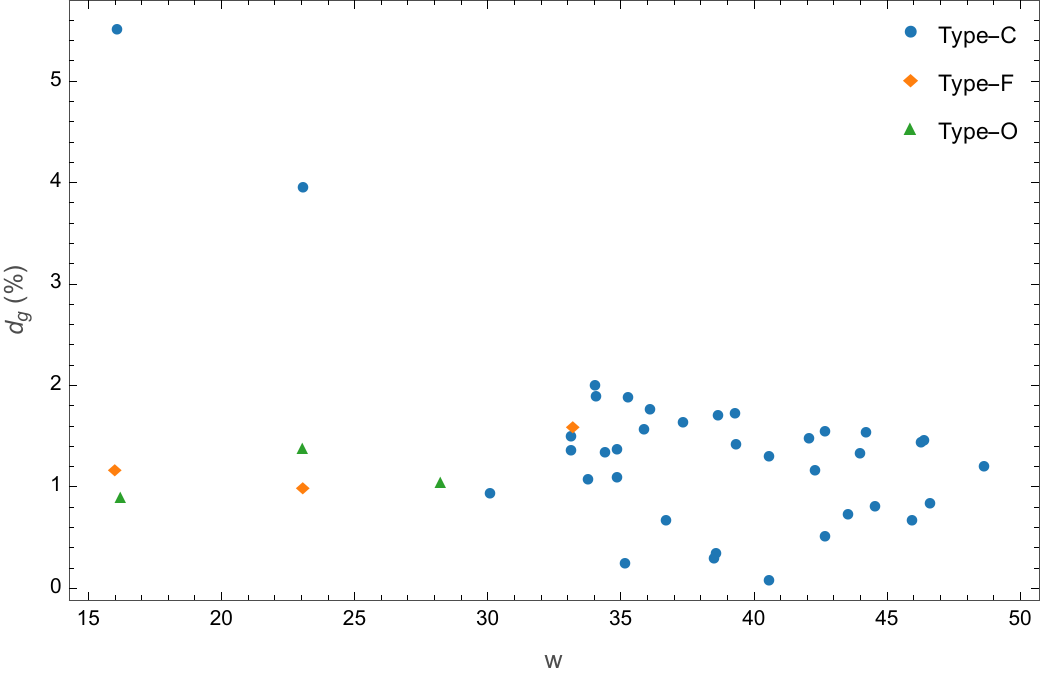}
  \end{minipage}
  \hfill
  \begin{minipage}[c]{0.49\linewidth}
    \centering
    \includegraphics[width=\linewidth]{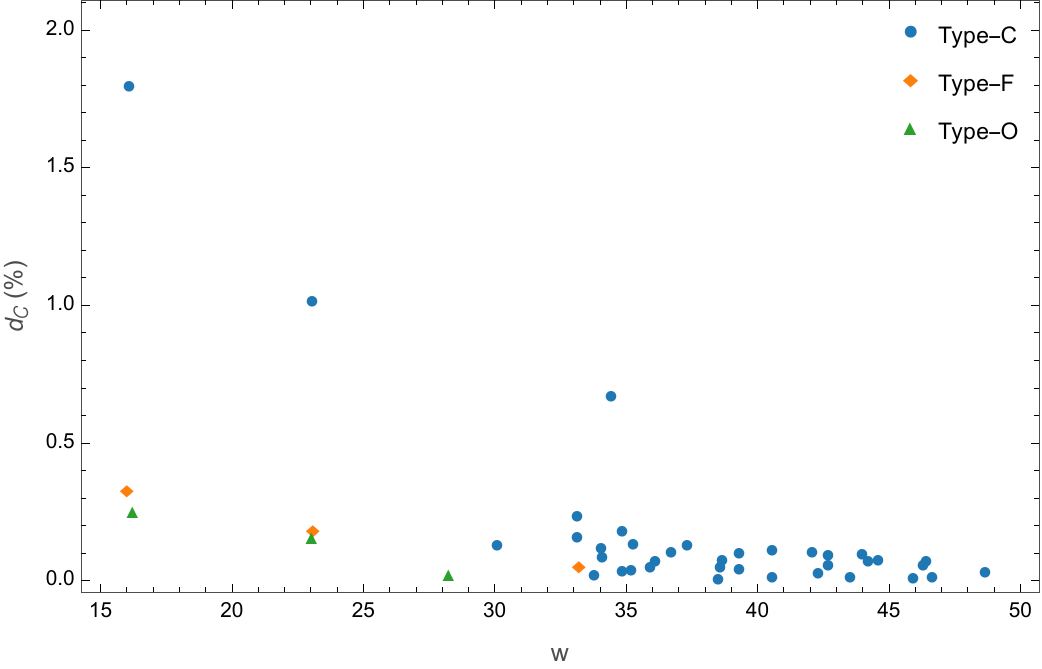}
  \end{minipage}
  \caption{Relative deviation, $d$, for Type-I perturbations, considering the analytical formula \eqref{deltaA}. The left panel is the relative deviation in terms of $g$ and the right panel in terms of $\mathcal{C}$. Here, one can clearly see that for all profiles considered, the relative deviation $d_\mathcal{C}<2\%$.}
    \label{fig:relatdifTI}
\end{figure}

As we already mentioned, we shall work instead with $g$, which is more convenient for Type-II perturbations. Notice, however, that because the relation between $\mathcal{C}$ and $g$ is quadratic \eqref{comp}, the error in terms of $g$ is generically higher than that of $\mathcal{C}$ (see Fig.\ref{fig:relatdifTI}).

\begin{figure}[h!]
    \centering
    \includegraphics[width=0.6\linewidth]{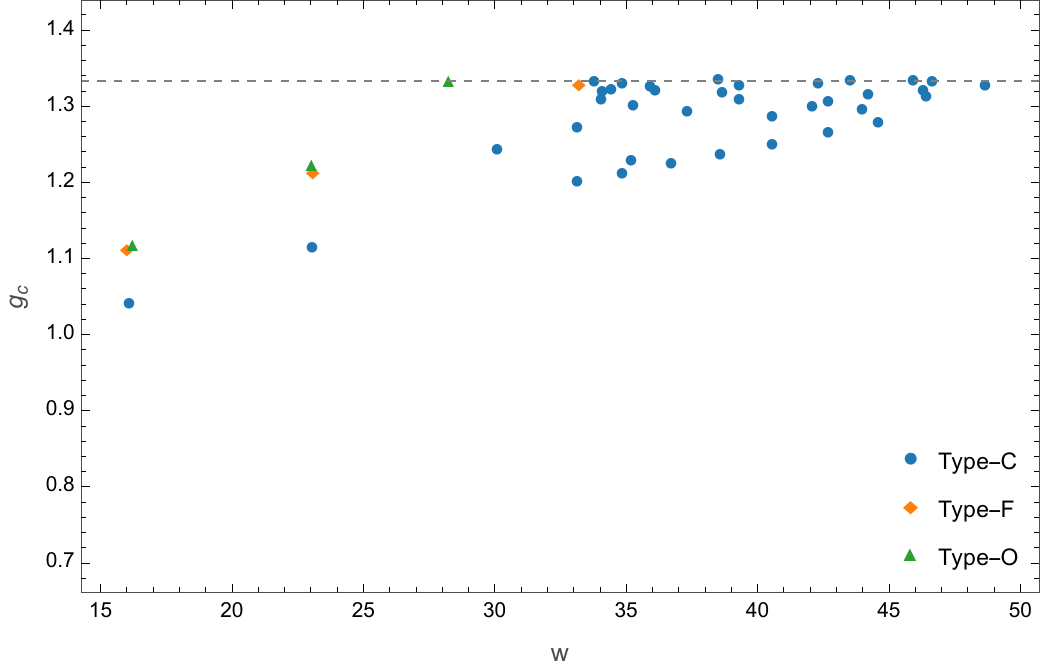}
    \caption{Numerical threshold values, in  terms of $g$, for Type-I perturbations.} 
    \label{fig:gcTI}
\end{figure}

In Fig. \ref{fig:gcTI}, we selected, from Fig.\ref{fig:gcAll}, all Type-I configurations.

There, the relative deviation between the analytical threshold ($\delta_A$) and the numerical one ($\delta_N$), $d\equiv|\delta_N-\delta_A|/\delta_N$, is $\sim 5\%$ (which reduces to $< 2\%$ for $\mathcal{C}_c$) for values of $w\sim 16$ and decreases to $\sim 1.8\%$, as $w$ increases ($< 0.15\%$ for $\mathcal{C}_c$), c.f. Fig.\ref{fig:relatdifTI}.

\subsection{Type-II}
Type-II perturbations require a higher threshold with respect to those of Type-I, as there the core would not help for the collapse of the external (higher density) shell. Illustrative examples are given in Fig.\ref{fig:Rall}, where Type-II is obtained by lowering the core from a Type-C(I).

\begin{figure}[h!]
  \centering
  \begin{minipage}[t]{0.48\linewidth}
    \centering
    \includegraphics[width=\linewidth]{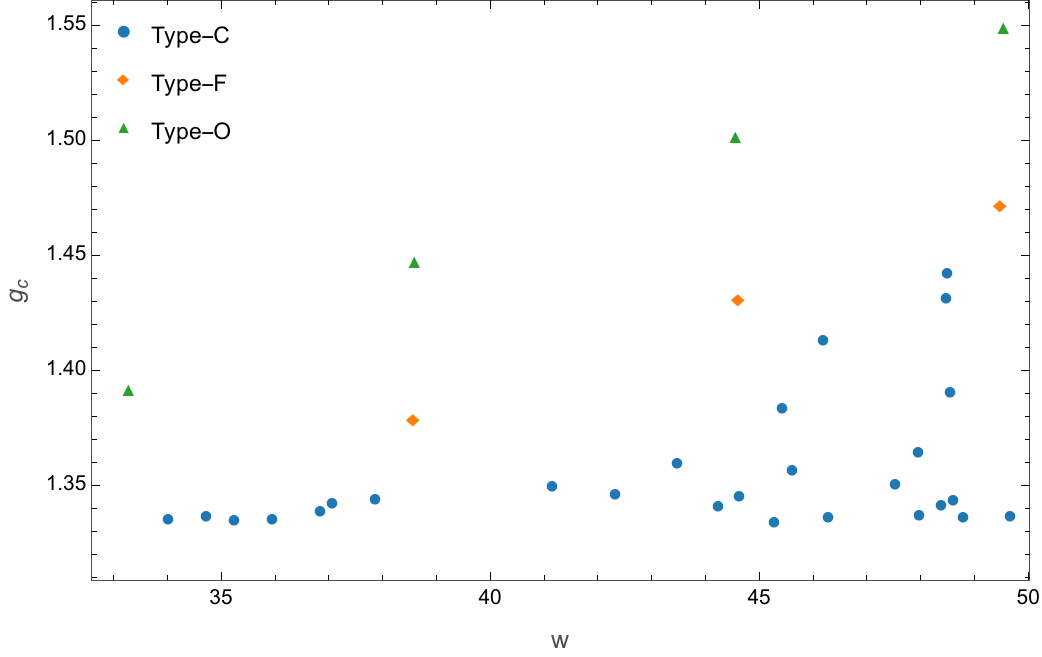}
  \end{minipage}
  \hfill
  \begin{minipage}[t]{0.48\linewidth}
    \centering
    \includegraphics[width=0.99\linewidth]{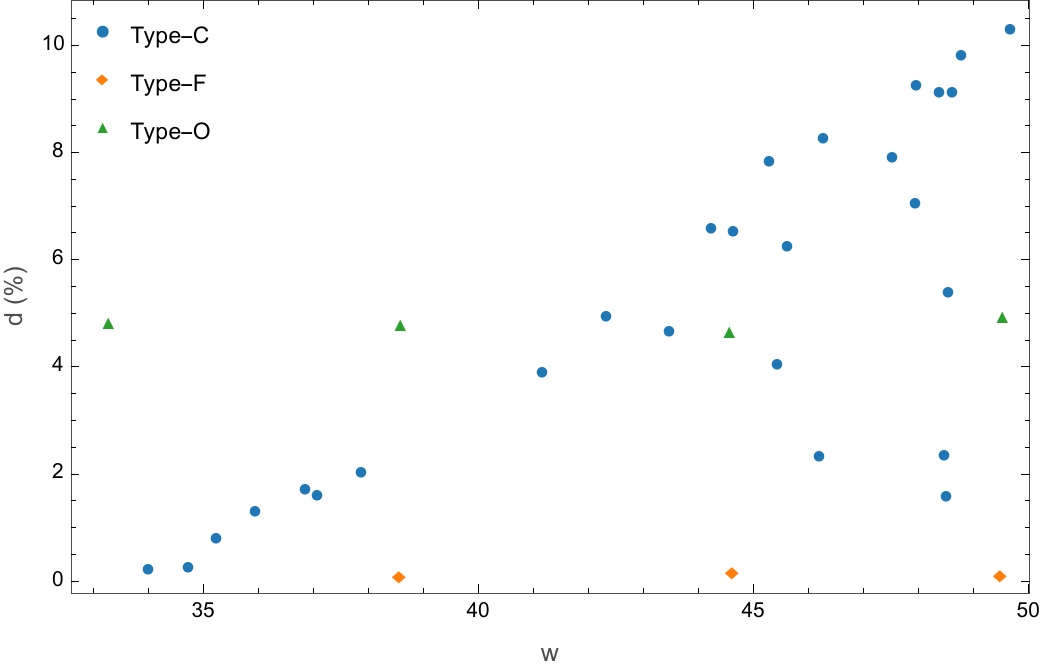}
  \end{minipage}
  \caption{The left panel is the numerical threshold, in terms of $g$, for Type-II perturbations. The right panel is the relative deviation, $d$, in terms of $g$ for Type-II perturbations using \eqref{gAalbert}.}
    \label{fig:thresholdsII}
\end{figure}

In \cite{w} it was shown that, at order $w\sim 30$ all the initial profiles considered (that can be checked to be of Type-F), at threshold, would produce Type-II black holes. In Fig.\ref{fig:gcTI}, we have explicitly shown that this is not always the case for Type-C, while it is true for Type-F and O. Nevertheless, there is also a distinction between those two. Considering the fitting formula proposed in \cite{w}, i.e.
\begin{equation}
    g_{c}=aw^b \hspace{1cm} (a\approx0.51962,\ b\approx0.26687), \label{gAalbert}
\end{equation}
for $w\gtrsim30$, one can see that it would only work for Type-F. Testing \eqref{gAalbert} for flat profiles \eqref{typeF} leads to a relative deviation with respect to the numeric of less than $0.3\%$. However, for Type-O, this analytical formula does not fit so well the numerical values, leading to a relative deviation of $\sim5\%$. For those Type-C profiles that cross to the Type-II region, the proposed analytical formula gives us a full range of values for similar values of $w$, from $d\sim 10\%$, which clearly largely deviate from the analytical formula, to $d\sim 2\%$ (see Fig.\ref{fig:thresholdsII}). 

For $w\lesssim30$, instead, the thresholds of all three types of profiles follow the analytical formula \eqref{deltaA} reasonably well. The reason is that, in this case, the maximum of $\R$ (or equivalently the amplitude of the collapsing shell) is broad and acts as a closed FRW while the core becomes insignificant. As $w$ increases, the shell gets thinner (and it cannot be considered a closed FRW anymore), making the core more and more relevant, that is when the trichotomy plays a role. This can clearly be observed in Fig.\ref{fig:gcAll}.

\section{Transition between closed and flat perturbations}

As we have shown, Type-C perturbations are characterised by constant strictly positive core values for $\R(x)$ or $\R(r)$ (Fig.\ref{fig:Rall}). 

One may ask what happens if this constant value decreases approaching zero, if it would still behave as Type-C or rather as a Type-F\footnote{We thank Jaume Garriga for posing this question.}.

As we have argued, the trichotomy appears as the shell gets thinner, and the core plays a role. Then, if the core corresponds to a closed FRW, it helps the collapse of the perturbation.
In this case, a ``low'' threshold is in order, leading to a Type-I BH. On the other hand, in the flat case, the core does not help the collapse, implying a higher threshold leading to Type-II BHs.

Because of the central importance of the size of the collapsing shell, we propose a new dimensionless parameter:
\begin{eqnarray}
    w_\R\equiv-r^2\frac{\partial^2_r\R(r)}{\R(r)}\Big|_{r=r_{p}}\ .
\end{eqnarray}

\begin{figure}[h!]
  \centering
  \begin{minipage}[t]{0.48\linewidth}
    \centering
    \includegraphics[width=\linewidth]{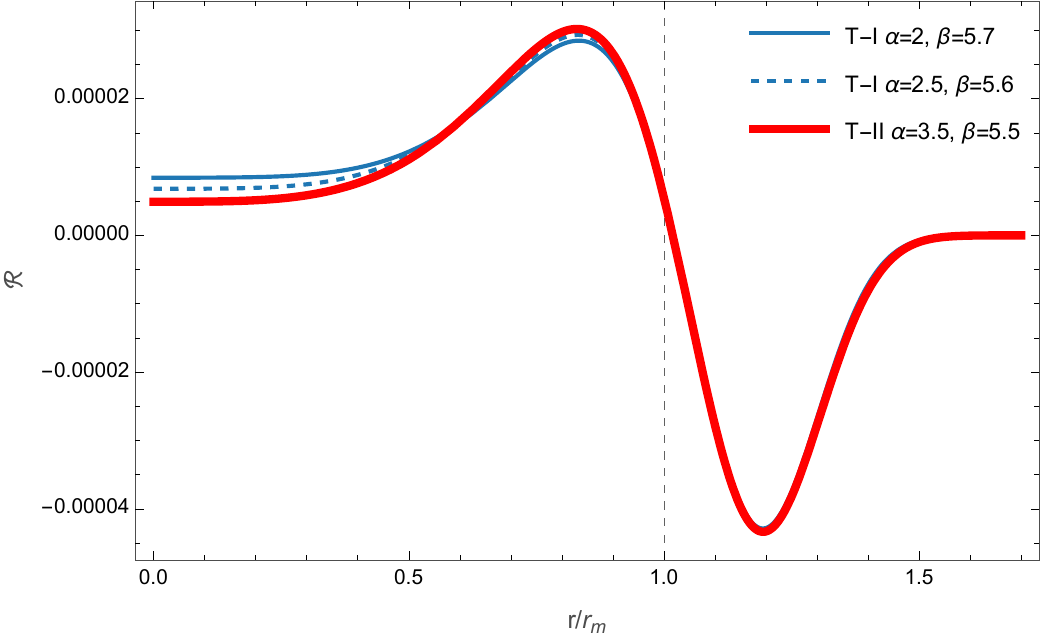}
  \end{minipage}
  \hfill
  \begin{minipage}[t]{0.48\linewidth}
    \centering
    \includegraphics[width=\linewidth]{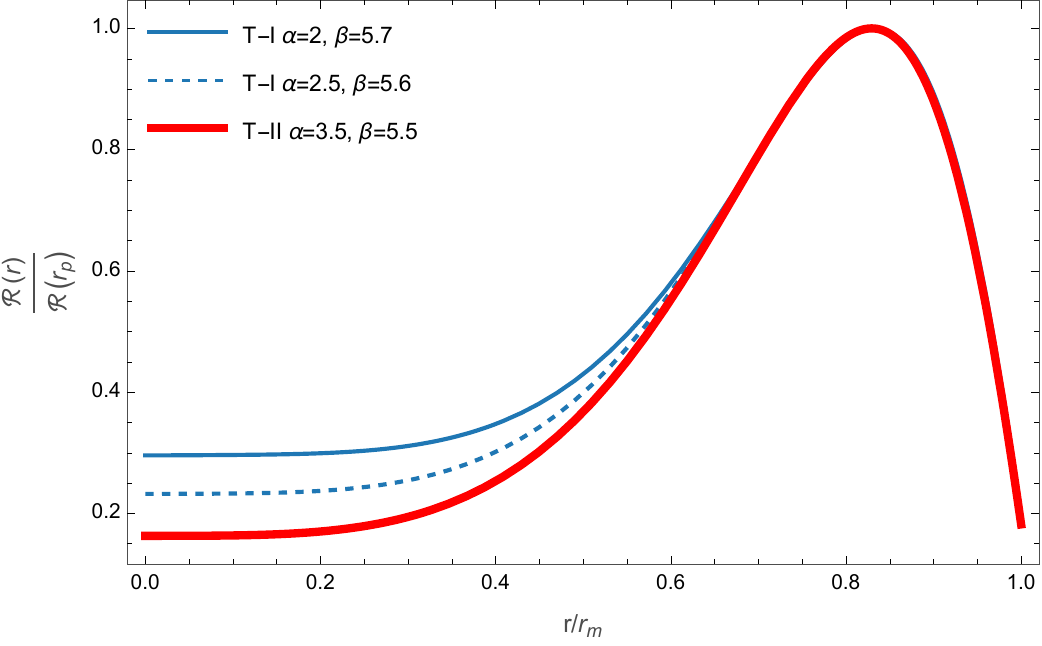}
  \end{minipage}
   \caption{The left panel is $\R$ for three different Type-C profiles, all of them corresponding to $w_\R\approx 27$. The first two profiles correspond to Type-I with different core values, while the last one corresponds to Type-II. The dashed vertical line is $r=r_m$. To make the comparison between the core and the peak of $\R$ clearer, in the right panel, we plot $\frac{\R(r)}{\R(r_p)}$ up to $r_m$.}
    \label{fig:wR27}
\end{figure}

In Fig.\ref{fig:wR27}, we can see that, fixing $w_\R$ we can go to a Type-II by lowering the core amplitude from a Type-I configuration with closed core (which we call Type-C(I)). Once the perturbation becomes Type-II, which requires a higher threshold, the core effect is negligible, and hence, it effectively acts as a flat one, we call this Type-C(II) configuration. 

As we are going to argue, the core plays an important role if it fulfils the following two conditions for $\R(r)$:
\begin{itemize}
    \item The amplitude of the core, $\R(0)$, is large enough with respect to the peak amplitude, $\R(r_p)$,.
    \item The peak is sharp enough, meaning $w_\R$ is large enough. 
\end{itemize}

\begin{figure}[h!]
    \centering
    \includegraphics[width=0.6\linewidth]{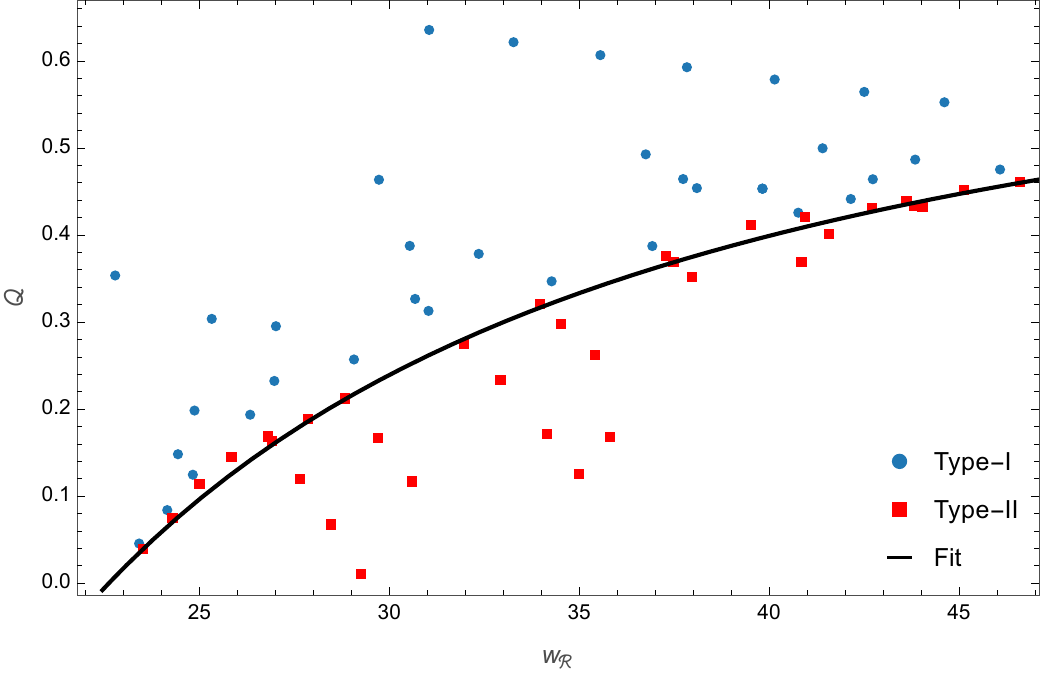}
    \caption{Values for ${\cal Q}\equiv\frac{\R(0)}{\R(r_{p})}\Big|_c$ for different values of $w_{\cal R}$. The blue points correspond to Type-I perturbations, while the red ones to Type-II. The line corresponds to a tentative transition region (Eq.\eqref{ratiofit}) that depends on the value of $w_{\cal R}$.All the data points correspond to the different profiles shown in Appendix \ref{appendix:parameters}.}
    \label{fig:wR}
\end{figure}

The first condition is actually connected to the second one and can be tested by computing the ratio
\begin{equation}
    {\cal Q}\equiv\frac{\R(0)}{\R(r_{p})}\Big|_c,
\end{equation}

in terms of the dimensionless curvature of the peak of $\R$, $w_\R$.
The sub-index $c$ in $\cal Q$ means that we are evaluating this quantity at the threshold. As we can see in Fig. \ref{fig:wR}, perturbations with a sharper peak require an even larger core in order to be of Type-I. The result is a non-constant transition region between Type-I and Type-II perturbations. This region is clearly parametrised by $w_{\cal R}$ as shown in the same figure.
 
If the peak is broad enough (i.e. $w_{\cal R}$ small), we can consider it as a closed FRW, which will help the collapse. Then, the core behaviour is not noticeable. That is the reason why, for small values of $w_{\cal R}$ the three types of perturbations are Type-I. On the other hand, for higher values of $w_{\cal R}$, the peak gets thinner and loses relevance with respect to the core.

The transition region between Type-I and II black holes obtained in Fig.\ref{fig:wR} follows the tentative formula:
\begin{equation}
    {\cal Q}=0.70-\frac{70.78}{(w_{\cal R})^{1.48}}.\label{ratiofit}
\end{equation}

\begin{figure}[h!]
    \centering
    \includegraphics[width=0.6\linewidth]{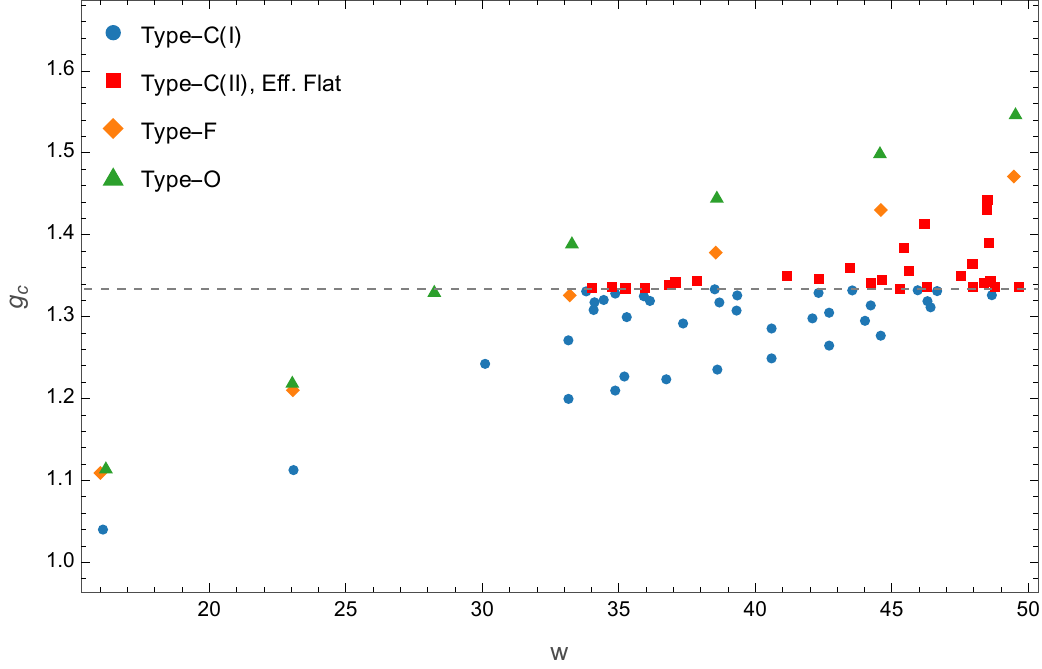}
    \caption{Numerical threshold, in terms of $g$ as a function of $w$, for the three types of perturbations, including the Type-C(II) profiles that behave effectively as flat. The dashed line corresponds to $g=\frac{4}{3}$. All the data points correspond to the different profiles shown in Appendix \ref{appendix:parameters}.}
    \label{fig:gc4-w}
\end{figure}

Once we have identified the region of Type-C perturbations with a negligible core, we can interpret the Type-C(II) points as effective Type-F ones (those below the curve $\cal Q$, c.f. Fig.\ref{fig:gc4-w}).

For $w_\R\lesssim 22$ ($w\lesssim30$), where ${\cal Q}=0$, all configurations lead to Type-I black holes while, for $w_\R\xrightarrow[]{}\infty$, our fitting formula for $\cal Q$ indicates that, in order to have Type-I configuration one needs a core amplitude to be at least $\sim 70\%$ of the peak one.  

\begin{figure}[h!]
    \centering
    \includegraphics[width=0.6\linewidth]{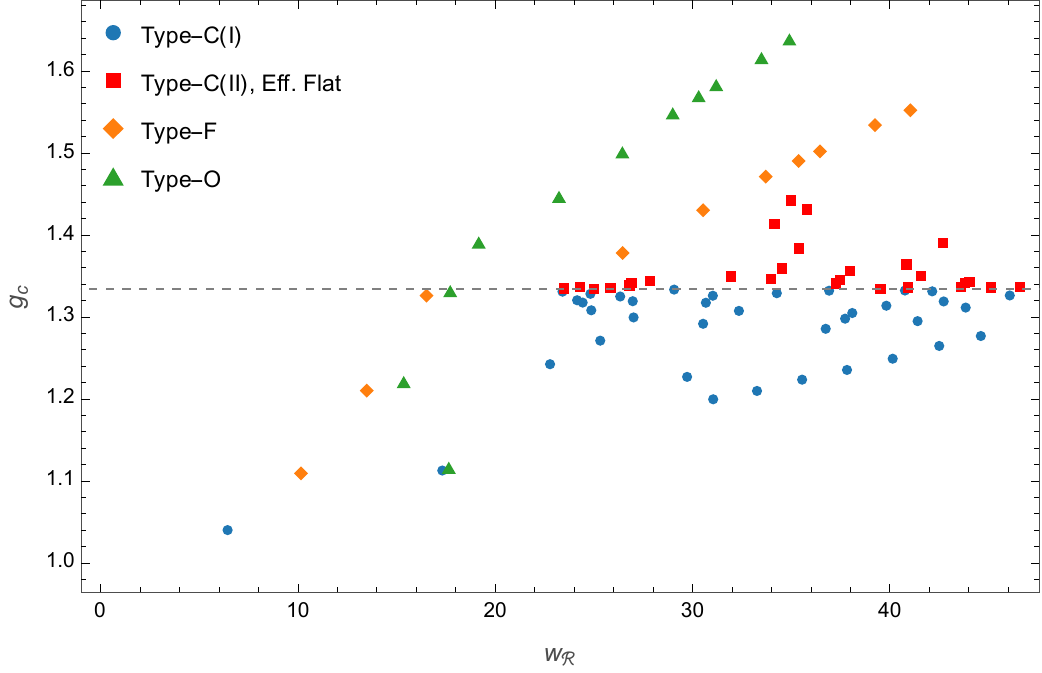}
    \caption{Numerical threshold, in terms of $g$ as a function of $w_{\cal R}$, for the three types of perturbations, including the effective flat profiles. The dashed line corresponds to $g=\frac{4}{3}.$}
    \label{fig:gc3-wR}
\end{figure}

In Fig.\ref{fig:gc4-w} and \ref{fig:gc3-wR}, one can see that $g_c$ in terms of $w_{\cal R}$ or $w$ is not as informative as $\cal Q$. The reason is that in $g$ the information about the core and peak of $\R$ is highly hidden. That's why we suggest that the three-dimensional curvature should be more properly used, rather than $g(r)$ or $\mathcal{C}(r)$, for threshold configurations.

\section{Initial conditions statistics}
PBHs are formed by rarely large smoothed over-densities on a radius $r$ \cite{sheth}. The threshold is then typically given in terms of the compaction function, which, if the perturbation is spherical, is the combination \eqref{comp} of the statistical variable \cite{sheth}
\begin{equation}
    g(r,\vec{x}_0)=-\frac{4}{9}\int \frac{d^3k}{(2\pi)^3}e^{i\vec{k}\cdot\vec{x_0}}(kr)^2W(kr)\zeta_k,
\end{equation}
where $r=|\vec{x}-\vec{x}_0|$ and $\zeta_k$, assumed to be independent and Gaussian-distributed, are the Fourier modes of the curvature perturbations in synchronous gauge and at leading order in gradient expansion. Note that here we have not yet assumed any spherical symmetry and
\begin{equation}
    W(y)=3\frac{\sin(y)-y\cos(y)}{y^3}\ ,
\end{equation}
is the Fourier transformation of the Heaviside theta function. For spherically symmetric configurations of the over-density, as we already pointed out, $g(r,\vec{x}_0)= a_i H_i^2 r^2 \delta_r$, and the combination \eqref{comp} indeed corresponds to the effectively asymptotically flat ``gravitational potential'' of the perturbation.

Typically, one expects that a black hole is formed by a large over-density (peak) at the point $\vec{x}=\vec{x}_0$, which statistically implies an almost spherical symmetry around $\vec{x}_0$ \cite{bbks}. In this case, without any loss of generality, one can fix $\vec{x}_0$ as the origin of the spherical symmetry.

Rewriting \eqref{Rr} in terms of $g$ and expanding it near the origin
one finds
\begin{equation}
    \frac{\R(r)}{H_i^2}=\left.\frac{3}{8}\frac{e^{-2 \zeta (r)} \left(8g(r)-3 g(r)^2+8rg'(r)\right)}{r^2{H_i^2}}\right|_{r\xrightarrow{}0}\approx 3e^{-2 \mu}\lim_{r\rightarrow 0}\frac{g(r)}{r^2H_i^2}=3e^{-2 \mu}\lim_{r\rightarrow 0}\delta_r\simeq \delta_0\ ,
\end{equation}
where $\delta_0$ is the central over-density\footnote{Note that, at the origin, the smoothed linear over-density is proportional to the non-linear over-density.} and we have used the empirical fact that $3e^{-2\mu}\sim\mathcal{O}(1)$ at threshold. 

We then already see that in the case of Type-F configuration, where $\delta_0=0$, instead of a peak of the over-density we are in the presence of a {\it trough} (a minimum). Thus, statistically, Type-C configurations (or effectively Type-F) are hardly spherical \cite{michiro}. In this case, the threshold for PBH formation will be larger than the one discussed above, implying an extra statistical suppression for their realisation with respect to the Type-C case. On the other hand, as we are going to see, the Type-C(II) core is less statistically suppressed with respect to the Type-C(I) one.

The estimation of which core type is finally most relevant for the PBHs abundance would then require a better statistical and numerical understanding of non-spherically symmetric configurations, although the numerical evolution of those is already an active field of research \cite{nonsph}. Nevertheless, in certain limiting configurations of the power spectrum, the selection of the specific initial conditions might be predicted without this knowledge, as we shall see.

The variable $\delta_0$ \footnote{Earlier statistics on PBHs abundances were constructed on this variable, e.g. \cite{mg}.}, as $g$, is multi-Gaussian with variance \cite{jacopo},
\begin{equation}
    \sigma_{\delta_0}^2= \frac{4}{81 H_i^4} \int dk\ k^3\mathcal{P}(k)\ ,
\end{equation}
where $\mathcal{P}(k)$ is the dimensionless power spectrum of the curvature perturbations $\zeta_k$.

\begin{figure}[h!]
    \centering
    \includegraphics[width=0.65\linewidth]{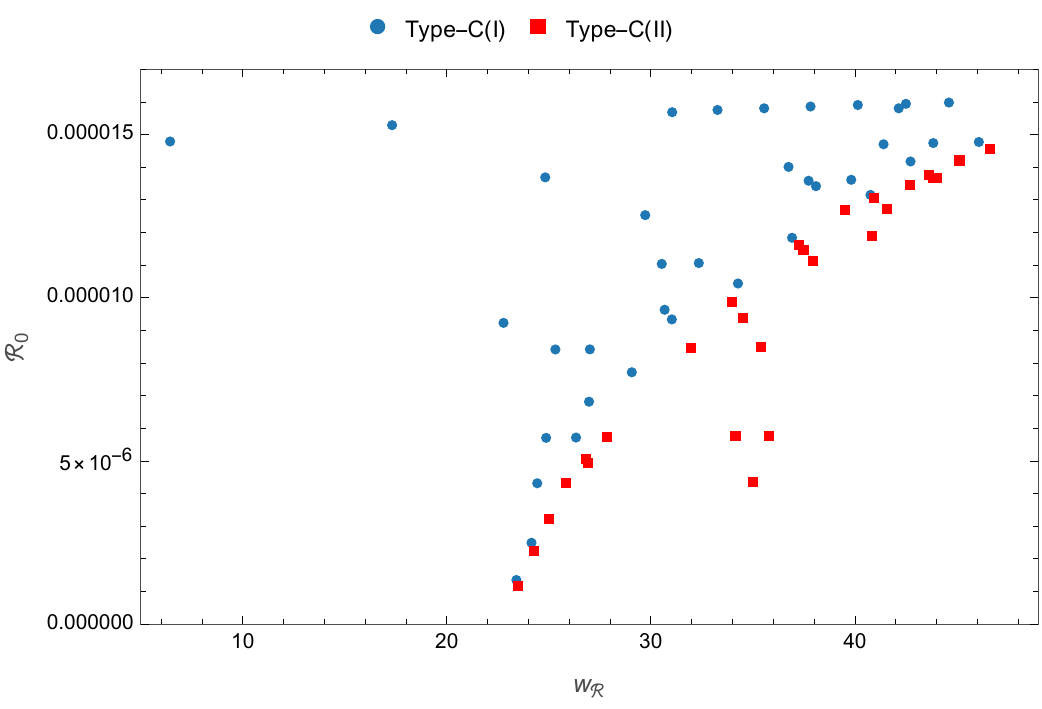}
    \caption{Values of $\R(r)$ at the centre for Type-C perturbations.}
    \label{fig:R0}
\end{figure}

For illustrative purposes, we will consider the two typical examples of a very narrow and a very broad $\mathcal{P}(k)$:
\begin{itemize}
    \item Narrow distribution, $\mathcal{P}(k)=\mathcal{A}_sk_p\delta(k-k_p)$.
    \begin{equation}
        \sigma_{\delta_0}^2= \frac{4}{81 H_i^4}\mathcal{A}_s k_p^4.
    \end{equation}
    \item Broad distribution, $\mathcal{P}(k)=\mathcal{A}_s\theta(k-k_{IR})\theta(k_{UV}-k)$, with $k_{UV}\gg k_{IR}$.
    \begin{equation}
        \sigma_{\delta_0}^2= \frac{1}{81 H_i^4}\mathcal{A}_s k_{UV}^4.
    \end{equation}
\end{itemize}
In Fig.\ref{fig:R0} we see that at threshold, the core values for Type-C(I) are of $\sim \mathcal{O}(10^{-5}-10^{-6})$ for all the cases studied.

Let us now consider that the peak of the PBH mass spectrum is associated to a specific Fourier mode $k_\bullet$. The order of magnitude of the PBH mass ($M_\bullet$) is \cite{musco} $M_{\bullet}\sim \frac{M_p^2}{H_k}$, where $H_k$ is the Hubble scale when the mode $k$ crossed the horizon. One then finds
\begin{eqnarray}
    k_\bullet\sim M_p\sqrt{\frac{H_i}{M_\bullet}}\ .
\end{eqnarray}

We start by assuming that $k_{UV}$ and $k_{p}$ are of order $k_\bullet$, which is what is expected for a very sharp power spectrum \cite{jacopo} and (optimistically \cite{jacopo}), $\mathcal{A}_s\sim 10^{-1}$, then 
\begin{equation}
\sigma^2_{\delta_0}\sim 10^{-2}\frac{M_p^2}{M_\bullet^2}\ .
\end{equation}
To maximise $\sigma^2_{\delta_0}$, we consider the lowest asteroid mass PBH (unconstrained to be the whole of dark matter \cite{Carr}). In this case  $M_{\bullet}\sim10^{41}\text{ GeV}$ and so
\begin{equation}
\sigma^2_{\delta_0}\sim 10^{-46}.
\end{equation}
Then,
\begin{equation}
    \sqrt{\sigma_{\delta_0}^2}\ll \delta_0(\text{Type-C(I)}) \sim 10^{-5} ,
\end{equation}
implying a statistical suppression of Type-C(I) profiles with respect to Type-F, since one should go almost $10^{18}$ variances away from the central value to obtain Type-C(I) perturbations.

For a very broad power spectrum, however, the peak of the mass distribution might instead be related to the IR scale, as it would be in the presence of only Type-I black holes \cite{jacopo}. In that case, for Type-C(I) not to be statistically suppressed, one would need $k_{UV}/ k_{IR}>10^{9}$ to obtain $ \sqrt{\sigma_{\delta_0}^2}\sim \delta_0(\text{Type-C(I)})$. Then, the analysis of \cite{jacopo} would follow since Type-C(I), given their lower threshold, would dominate the abundance of PBHs. 

We note that in \cite{NANOGrav}, such a large separation of scales were considered when interpreting the NANOGrav signal \cite{nano} as a second-order Gravitational Waves background sourced by large curvature perturbations.

\section{Conclusions}

According to the separate Universe approach, a local super-horizon patch (the core) of a perturbed Universe is described by a FRW geometry with a certain three-dimensional curvature. At larger scales, instead when the inhomogeneities build up, the typical configuration related to a PBH formation is a shell with an over-density peak. The formation of a BH happens whenever the amplitude of this shell is high enough for the gravitational collapse to fight against the Universe dilution. If the core is a closed Universe (Type-C) configuration, it helps the collapse of the shell, otherwise, either it fights against it (Type-O(pen)), or it is irrelevant (Type-F(lat)). Therefore, for Type-F and O configurations, the required amplitude of the energy density to form a black hole (the threshold) is bound to be higher than that of Type-C. In this sense, there is a trichotomy of possible initial conditions, each one having a different threshold for BH formation depending upon the three-curvature peak and core amplitudes. In turn, this shows that the relevant variable for understanding BH formation in an expanding Universe is rather the non-linear over-density instead of the compaction function or the {\it linear} over-density, as thought earlier. 

This trichotomy is, however, only relevant if the core is extended enough to interfere with the shell. This empirically happens if the dimensionless curvature of the over-dense shell $w_\R\equiv -r^2 \frac{\partial_r^2\R}{\R}$ calculated at the peak $r=r_p$, is higher than $\sim {\cal O}(17)$ ($\sim {\cal O}(30)$ for $w$), which is when the first point of Type-O crosses the Type-II region. For broader cases, the type of black hole formed is always of Type-I. However, when the core becomes relevant, only Type-C configuration with large enough core amplitudes may generate PBHs of Type-I. In all other cases, Type-II black holes will always be formed. For Type-C, the separatrix between Type-I and Type-II core amplitudes has been empirically found in eq. \eqref{ratiofit}, and depends upon the ratio of the core amplitude with respect to the peak one, being parametrised by $w_\R$. For very sharp shells, we found that the core amplitude must be at least about $70\%$ the peak one in order to form Type-I black holes. 

A more extensive and complete analysis should be done considering different perturbation profiles, $\zeta(r)$, which we leave for future work.

Finally, we argued that, at least for Gaussian distributed curvature perturbations, shapes with zero curvature cores (Type-F in spherical configuration) are favourable. However, whether black holes will be more likely related to these shapes is a different question, which can only be answered with the statistics of thresholds. For example, Type-F configurations are also likely to generate a highly non-spherical initial condition for the non-linear over-density (note that this would be statistically different from the linear one), making gravitational collapse more difficult than in the spherical case. The reason is that a sharp shell can be thought of as generated by the interference of an infinite number of peaks, which would distort the spherical symmetry. For highly non-spherical configurations, a large part of the energy density of the over-dense shell would be dissipated away in the form of gravitational waves, thus a larger threshold. 

For a very broad power spectrum, however, which might be related to the NanoGrav gravitational waves observations \cite{NANOGrav}, the statistical preference of Type-F with respect to Type-C(I) becomes negligible, in that case $ \sqrt{\sigma_{\delta_0}^2}\sim \delta_0(\text{Type-C(I)})$. In this case, the higher threshold necessary to form Type-F black holes would favour Type-C(I) for the PBHs abundance. This, in turn, would imply a black hole's mass spectrum peaked at the Infra-Red scale of the power spectrum \cite{jacopo}, instead of at the Ultra-Violet scales, as typically assumed.

Whether for a generic power spectrum, the abundance of PBHs is finally dominated by Type-I black holes, as suggested in \cite{jacopo}, or not, is however still an open question which, to be answered, requires further (physical and statistical) knowledge of the non-spherical gravitational collapse. This interesting problem is left for future studies.

\begin{acknowledgments}
The authors are grateful to Albert Escriv\`a, Jacopo Fumagalli, Jaume Garriga, Cristian Joana and Ravi Sheth for discussions on thresholds, profiles and Types-I/II initial conditions. 
The research of CG is supported by the
grant PID2022-136224NB-C22, funded by MCIN\allowbreak/\allowbreak AEI\allowbreak/10.13039\allowbreak/501100011033\allowbreak/\allowbreak FEDER,
UE, by the grant\allowbreak/ 2021-SGR00872 and CEX2024-001451-M funded by MICIU/AEI/10.13039/501100011033. 
\end{acknowledgments}

\appendix
\section{Profile parameters}\label{appendix:parameters}
Here, we provide all the parameters used for each profile shown. 

\begin{table}[h!]
\begin{tabular}{|c|c|c|c|c|c||c|c|c|c|c|c|}
\hline
$\mu$& $w$   & $w_{\cal R}$ & $\lambda$ & $\alpha$ & $\beta$ & $\mu$& $w$   & $w_{\cal R}$ & $\lambda$ & $\alpha$ & $\beta$ \\ \hline\hline
0.678&16.11 & 6.45  & 0.939     & 1        & 5.53    & 0.605&38.69 & 30.68 & 0.983     & 1.74     & 6       \\ \hline
0.671&23.09 & 17.33 & 0.953     & 1        & 6.075   & 0.616&39.32 & 32.36 & 0.981     & 1.5      & 6.2     \\ \hline
0.613&30.11 & 22.79 & 0.981     & 1.8      & 5.5     & 0.602&39.34 & 31.04 & 0.984     & 1.8      & 6       \\ \hline
0.600&33.16 & 25.33 & 0.984     & 2        & 5.7     & 0.658 &40.60 & 40.15 & 0.969     & 1        & 7.1     \\ \hline
0.665&33.16 & 31.05 & 0.964     & 1        & 6.7     & 0.634&40.60 & 36.76 & 0.974     & 1.1542   & 6.7     \\ \hline
0.551&33.81 & 23.42 & 0.998     & 13       & 5.1     & 0.636&42.10 & 37.74 & 0.976     & 1.2      & 6.7     \\ \hline
0.581&34.08 & 24.88 & 0.990     & 3        & 5.4     & 0.610&42.32 & 34.28 & 0.983     & 1.6      & 6.2     \\ \hline
0.571&34.12 & 24.44 & 0.993     & 4        & 5.3     & 0.634&42.71 & 38.10 & 0.976     & 1.2175   & 6.7     \\ \hline
0.558&34.45 & 24.16 & 0.996     & 7        & 5.2     & 0.658&42.71 & 42.51 & 0.970     & 1        & 7.2     \\ \hline
0.662&34.87 & 33.27 & 0.965     & 1        & 6.8     & 0.619&43.56 & 36.93 & 0.980     & 1.4      & 6.5     \\ \hline
0.565&34.88 & 24.83 & 0.994     & 4.7      & 5.3     & 0.645&44.02 & 41.41 & 0.974     & 1.1      & 6.99    \\ \hline
0.600&35.30 & 27.02 & 0.985     & 2        & 5.7     & 0.635&44.23 & 39.83 & 0.977     & 1.2      & 6.8     \\ \hline
0.633&35.21 & 29.73 & 0.975     & 1.3      & 6.2     & 0.657&44.60 & 44.62 & 0.971     & 1        & 7.29    \\ \hline
0.578&35.92 & 26.34 & 0.990     & 3        & 5.5     & 0.629&45.95 & 40.77 & 0.978     & 1.25     & 6.8     \\ \hline
0.586&36.15 & 26.98 & 0.988     & 2.5      & 5.6     & 0.638&46.31 & 42.74 & 0.976     & 1.15     & 6.99    \\ \hline
0.661&36.74 & 35.56 & 0.966     & 1        & 6.9     & 0.644&46.43 & 43.85 & 0.975     & 1.1      & 7.1     \\ \hline
0.618&37.36 & 30.54 & 0.980     & 1.5      & 6.1     & 0.634&46.67 & 42.15 & 0.978     & 1.2      & 6.91    \\ \hline
0.590&38.52 & 29.07 & 0.987     & 2.2      & 5.8     & 0.643&48.67 & 46.08 & 0.976     & 1.1      & 7.2     \\ \hline
0.659&38.62 & 37.84 & 0.968     & 1        & 7       &       &       &           &          &       &  \\ \hline
\end{tabular}
\caption{Parameters used for Type-C(I) profiles in Fig.\ref{fig:gcTI} and Fig.\ref{fig:relatdifTI} and the corresponding $w$ and $w_{\cal R}$.}
    \label{tab:parametersC}
\end{table}
\newpage
\begin{table}[h!]
  \centering
  \begin{minipage}[t]{0.45\linewidth}
    \centering
    \begin{tabular}{|c|c|c|c|c|}\hline
         Type&$\mu$&$w$&$w_{\cal R}$&$\beta$ \\ \hline\hline
         I&0.595&16.01&10.17&3.8\\ \hline
         I&0.565&23.10&13.50&4.3666\\ \hline
         I&0.540&33.21&16.53&5.005\\ \hline
         II&0.531&38.56&26.47&5.29\\ \hline
         II&0.522&44.61&30.55&5.585\\ \hline 
         II&0.517&49.48&33.72&5.8\\ \hline
    \end{tabular}
    \caption{Parameters used for Type-F profiles in Fig.\ref{fig:gcTI} and Fig.\ref{fig:thresholdsII} and the corresponding $w$ and $w_{\cal R}$.}
    \label{tab:parametersF}
  \end{minipage}
  \hfill
  \begin{minipage}[t]{0.45\linewidth}
    \centering
    \begin{tabular}{|c|c|c|c|c|c|c|}\hline
         Type&$\mu$&$w$&$w_{\cal R}$&$\lambda$& $\alpha$&$\beta$ \\ \hline\hline
         I&0.595&16.22&17.66&1.001&70&3.8\\ \hline
         I&0.558&23.05&15.37&1.004&10&4.27\\ \hline
         I&0.457&28.25&17.73&1.026&10&4.1\\ \hline
         II&0.444&33.29&19.17&1.022&\multirow{4}{*}{1}&4.4\\ \cline{1-5}\cline{7-7}
         II&0.433&38.60&23.24&1.019&&4.685\\ \cline{1-5}\cline{7-7}
         II&0.422&44.57&26.45&1.017&&4.98\\ \cline{1-5}\cline{7-7}
         II&0.417&49.54&29.00&1.016&&5.194\\ \hline
    \end{tabular}
    \caption{Parameters used for Type-O profiles in Fig.\ref{fig:gcTI} and Fig.\ref{fig:thresholdsII} and the corresponding $w$ and $w_{\cal R}$.}
    \label{tab:parametersO}
  \end{minipage}
\end{table}
\begin{table}[h!]
\begin{tabular}{|c|c|c|c|c|c||c|c|c|c|c|c|}
\hline
$\mu$&$w$   & $w_{\cal R}$ & $\lambda$ & $\alpha$ & $\beta$ & $\mu$&$w$   & $w_{\cal R}$ & $\lambda$ & $\alpha$ & $\beta$ \\ \hline
0.550&34.02 & 23.51 & 0.998     & 15       & 5.1     & 0.611&45.63 & 37.97 & 0.983     & 1.5      & 6.5     \\ \hline
0.556&34.74 & 24.29 & 0.996     & 7.8      & 5.2     & 0.566&46.21 & 34.16 & 0.992     & 3        & 6       \\ \hline
0.563&35.25 & 25.10 & 0.994     & 5.1      & 5.3     & 0.628&46.29 & 40.95 & 0.979     & 1.26     & 6.8     \\ \hline
0.568&35.96 & 25.85 & 0.993     & 4        & 5.4     & 0.624&47.55 & 41.58 & 0.980     & 1.3      & 6.8     \\ \hline
0.572&36.86 & 26.81 & 0.992     & 3.4      & 5.5     & 0.616&47.96 & 40.86 & 0.982     & 1.4      & 6.7     \\ \hline
0.571&37.08 & 26.91 & 0.992     & 3.5      & 5.5     & 0.633&47.98 & 43.62 & 0.978     & 1.19     & 6.99    \\ \hline
0.576&37.88 & 27.85 & 0.991     & 3        & 5.6     & 0.632&48.39 & 43.83 & 0.978     & 1.2      & 6.99    \\ \hline
0.593&41.17 & 31.97 & 0.987     & 2        & 6       & 0.564&48.48 & 35.82 & 0.993     & 3        & 6.1     \\ \hline
0.604&42.34 & 33.97 & 0.984     & 1.7      & 6.2     & 0.553&48.51 & 35.00 & 0.995     & 4        & 6       \\ \hline
0.598&43.48 & 34.53 & 0.985     & 1.8      & 6.2     & 0.663&48.56 & 42.71 & 0.978     & 1.2      & 6.9     \\ \hline
0.617&44.25 & 37.28 & 0.981     & 1.43     & 6.5     & 0.632&48.63 & 44.05 & 0.978     & 1.2      & 7       \\ \hline
0.615&44.65 & 37.49 & 0.982     & 1.45     & 6.5     & 0.637&48.80 & 45.14 & 0.977     & 1.15     & 7.1     \\ \hline
0.625&45.30 & 39.54 & 0.979     & 1.3      & 6.7     & 0.640&49.68 & 46.61 & 0.977     & 1.12     & 7.2     \\ \hline
0.589&45.44 & 35.42 & 0.988     & 2        & 6.2     &       &     &  &           &          &         \\ \hline
\end{tabular}
\caption{Parameters used for Type-C(II) profiles in Fig.\ref{fig:thresholdsII} and Fig.\ref{fig:gc3-wR} and the corresponding $w$ and $w_{\cal R}$.}
    \label{tab:parametersefflat}
\end{table}

\section{Hamiltonian constraint}\label{appendix:hamiltonian}
\begin{figure}[h!]
    \centering
    \includegraphics[width=0.6\linewidth]{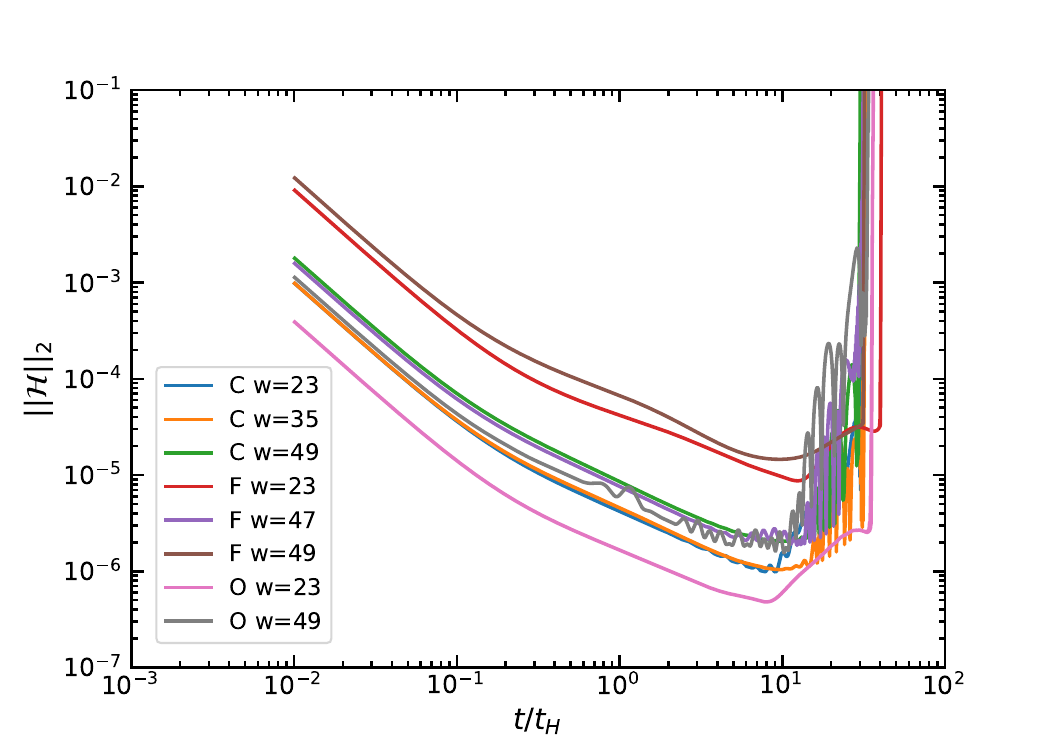}
    \caption{Numerical evolution of the Hamiltonian constraint for different profiles at threshold. The time of horizon crossing, when the perturbation re-enters the horizon, is $t_H=t_0\left(a_i r_m H_i\right)^2$, where $t_0=1$ for the simulations.}
    \label{fig:hamilt}
\end{figure}

To test the convergence of the simulations, we have analysed the evolution of the Hamiltonian constraint equation, (as in \cite{albert})
\begin{equation}
    \mathcal{H}=D_rM-4\pi\Gamma\rho R^2,
\end{equation}
\begin{equation}
    \left|\left|\mathcal{H}\right|\right|^2= \frac{1}{N_{cheb}}\sqrt{\sum_k \left|\frac{M'_{k}/R'_k}{4\pi\rho_k R_k^2}-1\right|^2}.
\end{equation}
The results for the evolution of the Hamiltonian constraint using the three profiles considered at threshold (Eq.(\ref{typeF})-(\ref{TypeO})), are shown in Fig.\ref{fig:hamilt}.
\\

In Fig.\ref{fig:hamilt}, one can see that around $t/t_H\sim10^1$, the Hamiltonian constraint evolution oscillates and diverges from the previous monotonic behaviour. That is because around that point the BH is formed, and then the singularity, hence the numerical evolution does not hold any more.

\end{document}